\documentclass[twocolumn,preprintnumbers,aps,showpacs,superscriptaddress,prl]{revtex4-1}

\usepackage{palatino}
\usepackage[latin1]{inputenc}
\usepackage{epsf}
\usepackage{amsmath,amssymb}
\usepackage{latexsym}
\usepackage{calc}
\usepackage{color}
\usepackage{shadow}
\usepackage{epsfig}
\usepackage{float}

\usepackage[pdftex, pdfstartview = {FitH}]{hyperref}

\newcommand{\ben}{\begin{equation}}
\newcommand{\een}{\end{equation}}
\newcommand{\bea}{\begin{eqnarray}}
\newcommand{\eea}{\end{eqnarray}}

\def\sss{\scriptscriptstyle\rm}

\def\xc{_{\sss XC}}

\begin{document}
\title{Exact time-dependent exchange-correlation potential in electron scattering processes}

\author{Yasumitsu Suzuki}
\affiliation{Department of Physics, Tokyo University of Science, 1-3 Kagurazaka, Shinjuku-ku, Tokyo 162-8601, Japan}  
\author{Lionel Lacombe}
\affiliation{Department of Physics and Astronomy, Hunter College and the Graduate Center of the City University of New York, 695 Park Avenue, New York, New York 10065, USA} 
\author{Kazuyuki Watanabe}
\affiliation{Department of Physics, Tokyo University of Science, 1-3 Kagurazaka, Shinjuku-ku, Tokyo 162-8601, Japan} 
\author{Neepa T. Maitra}
\affiliation{Department of Physics and Astronomy, Hunter College and the Graduate Center of the City University of New York, 695 Park Avenue, New York, New York 10065, USA} 

\date{\today}
%\pacs{31.15.-p, 31.50.-x, 82.20.Gk}

\begin{abstract}

We identify peak and valley structures in the exact exchange-correlation potential of time-dependent density functional
 theory that are crucial for time-resolved electron scattering in a model one-dimensional system. These structures are completely 
missed by adiabatic approximations which consequently significantly underestimate the scattering probability. 
A recently-proposed non-adiabatic approximation is shown to correctly capture the approach of the electron to the target 
when the initial Kohn-Sham state is chosen judiciously, and is more accurate than standard adiabatic functionals, 
but it ultimately fails to accurately capture reflection. These results may explain the underestimate of scattering 
probabilities in some recent studies on molecules and surfaces.
\end{abstract}

\maketitle

Electron scattering is one of the most fundamental processes 
in physics, chemistry, and biology. Electrons constantly collide with 
other electrons and nuclei in chemical reactions and physical processes, from molecular electronics~\cite{ME} to strong-field processes~\cite{SFP}. 
Some experimental techniques directly utilize electron scattering, such as
transmission or scanning electron microscopy~\cite{TEM,*SEM},  
to investigate surface atomic structures. Radiation damage caused by low-energy electron
 scattering from DNA highlights its relevance for biomolecules~\cite{DNA}. 
Despite the importance of electron scattering processes, its theoretical description 
 remains a most challenging problems: electron scattering is a highly-correlated many-body problem
  and generally requires treatment beyond perturbation theory.

Time-dependent density functional theory (TDDFT)~\cite{tddft1,*tddft2,*tddft3} is the most widely
used first-principles approach to study real-time many-electron dynamics. This is due to its favorable system-size scaling, which arises because it maps the
  correlated electron dynamics to that of the
non-interacting Kohn-Sham (KS) system, evolving in a single-particle potential. 
All many-body effects are hidden in the single-particle exchange-correlation (xc) potential, which in practice must be approximated.
TDDFT with approximate xc potentials has been successfully applied to interpret and predict 
electron dynamics in a range of situations~\cite{tddft4,*tddft5,*tddft6,*tddft7,*tddft8,*tddft9,*tddft10}, in addition to predicting 
linear response and spectra, which is the regime it is mostly known for. 
Indeed, it has been applied to compute elastic electron-atom scattering cross-sections
by means of linear-response theory~\cite{scat1,*scat2,*WMB05}, and
recently also applied to real-time non-perturbative calculations of proton-methane
 scattering~\cite{GWWZ14,*QSAC17} and of electron wavepacket scattering from graphene~\cite{scat3,*scat4, *scat5,*scat6}.

However, TDDFT with approximate xc potentials
fails to even qualitatively reproduce the true dynamics in some 
applications to non-linear time-resolved dynamics~\cite{fail1,*fail2,*fail3,*fail4}.
In principle, the exact xc potential $v_{\rm xc}$ at time $t$ functionally depends 
on the history of the density $n({\bf r},t'<t)$,
the initial interacting many-body state $\Psi_0$, and
the choice of the initial KS state $\Phi_0$.
 In reality, almost all calculations today use an adiabatic approximation that inputs the 
instantaneous density into a ground-state xc functional. 
Recent studies on exactly-solvable model systems~\cite{xc1,xc2,xc3,RG12}
reveal that large non-adiabatic features can appear in the exact xc potential
that are missing in the approximations. How accurate is TDDFT with  the currently available approximate xc potentials for 
electron scattering? Given the dearth of alternative practical ab initio methods for this problem
 and its relevance in a wide range of situations of interest today, it is crucial to assess the reliability of TDDFT approximations for scattering. 

To this end, we study a model system of electron-Hydrogen (e-H) scattering that can be solved numerically exactly, and 
show that the exact xc potential develops non-adiabatic peak and valley structures that are dominantly responsible for causing
 scattering. Standard functional approximations lack these structures and 
  severely underestimate the scattering probability. Although a recently-proposed non-adiabatic orbital-dependent
 functional significantly improves the dynamics in the approach of the electron to the target for a judiciously chosen initial KS state,
 it also ultimately fails to accurately scatter. We identify the term in the exact xc potential crucial for the scattering, 
as an explicit functional of the many-body density-matrix and KS orbitals. Our results stress the need to develop explicit
 density or orbital-functionals for this term, and suggest that, generally, adiabatic TDDFT tends to underestimate the scattering
 and energy-transfer in real systems~\cite{GWWZ14,*QSAC17, scat3,*scat4,*scat5}.
 We point out the
important difference in the performance of a given approximate
functional applied in the linear response regime to extract elastic
scattering cross-sections~\cite{scat1,*scat2,*WMB05} and application of the same functional in
simulating time-resolved wavepacket scattering.

The Hamiltonian of our one-dimensional two-electron model system reads (we use atomic units throughout unless otherwise stated): $
{\hat H}(x_1,x_2)=\sum_{i=1,2}\left(-\frac{1}{2}\frac{\partial^2}{\partial x_i^2}+v_{\rm ext}(x_i)   \right) +  W_{ee}(x_1,x_2)
$,
where  $W_{ee}(x_1,x_2)=\frac{1}{\sqrt{(x_1-x_2)^2+1}}$ is the soft-Coulomb interaction~\cite{soft}
and
 $v_{\rm ext}(x)=-\frac{1}{\sqrt{(x+10)^2+1}}$ is
the external potential that models the H atom located at $x=-10.0$ a.u.
The soft-Coulomb interaction has been used extensively in strong-field physics as well as in
density functional theory as it captures the essential physics of real atoms and molecules.

The initial interacting wavefunction is taken to be a spin-singlet, with spatial part
\ben
\Psi_0(x_1,x_2) = \frac{1}{ \sqrt{2}}\left(\phi_{\rm H}(x_1)\phi_{\rm WP}(x_2) +\phi_{\rm WP}(x_1)\phi_{\rm H}(x_2) \right)
\label{eqn:Psi}
\een
where $\phi_{\rm H}(x)$ is the ground-state hydrogen wavefunction. 
 $\phi_{\rm WP}(x)$ is an incident Gaussian wavepacket,
\ben
\begin{split}
\phi_{\rm WP}(x)=\left(2\alpha/\pi\right)^{\frac{1}{4}}e^{\left[-\alpha(x-x_0)^2+ip(x-x_0)\right]}
\end{split}\label{eqn: wp}
\een
with $\alpha=0.1$, 
representing an electron  at $x_0=10.0$ a.u. approaching the target atom with a momentum $p=-1.5$ a.u. in our first example.
The full time-dependent Schr\"odinger equation 
$i\partial_t\Psi(x_1,x_2,t)={\hat H}(x_1,x_2)\Psi(x_1,x_2,t)$
can be solved numerically exactly for this system, and we plot the resulting density,
 $n(x,t)=2\int |\Psi(x,x_2,t)|^2dx_2$, as the black lines in the upper panel for different time-slices 
in Fig.~\ref{fig:Fig1}~\footnote[1]{Movies of the dynamics are given in the Supplemental Material.}. 
\begin{figure}[h]
 \centering
 \includegraphics*[width=1.0\columnwidth]{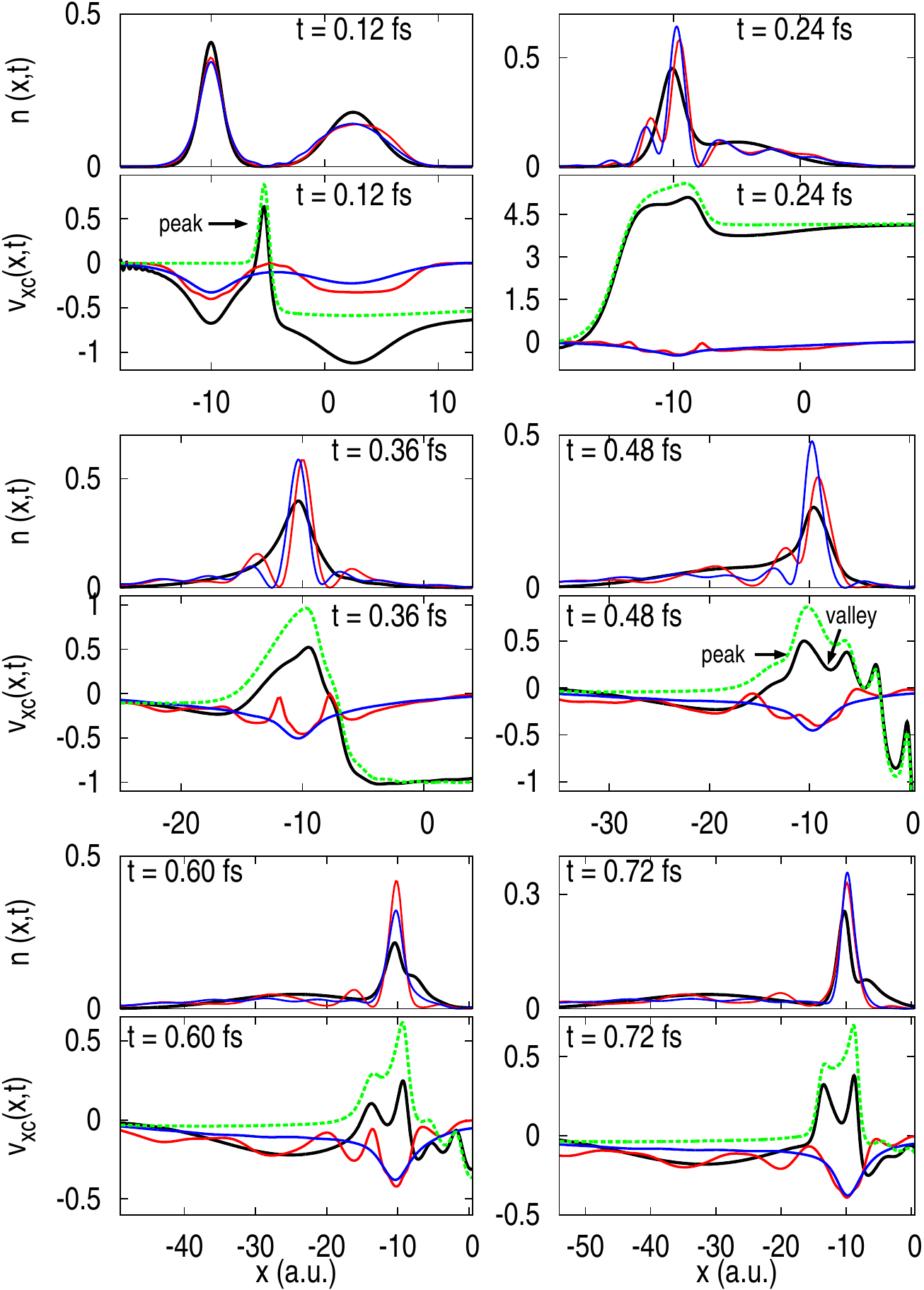}
 \caption{(color online). Snapshots of the exact electron density $n (x,t)$ in the e-H scattering model system
 (black solid line in the upper panel for each time slice).
Black line in the lower panel shows the exact time-dependent xc potential $v_{\rm xc}$ for the initial KS state $\Phi_0^{(1)}$ for each time slice.
The results of ALDA (red solid line) and $v^{\rm S}_{\rm xc}$ (blue solid line) are shown in each panel.
The kinetic component of the exact xc potential $v^{\rm T}_{\rm c}$ is also shown as green dotted line in the lower panels.
}
 \label{fig:Fig1}
\end{figure}
The incident wavepacket collides with the target electron at around 0.24 fs,
after which, a major part of the wavepacket is transmitted while some is reflected back. The black lines in top panels
 of Fig.~\ref{fig:Fig2} show the number of reflected (transmitted) electrons, $N_R$ ($N_T$), calculated by integrating $n(x,t)$
 over the region $x>-5.0$ a.u. ($x<-15.0$ a.u.); notice that the
incoming electron leaves the target partially ionized. At this incident momentum, the scattering is inelastic, evident in density
 oscillations in the target after the incoming electron has well passed
(evident in the movies in the supplementary information~\footnotemark[1]); later we shall consider the case of low-energy elastic scattering. 
Shown also in both these figures are the results from TDDFT approximations (red and blue lines), 
neither of which yield reflection nor the dynamics correctly.
 To understand why, we now consider the xc potential 
for an exact TDDFT calculation and compare this with the approximate potentials.

\begin{figure}[h]
 \centering
\includegraphics*[width=1.0\columnwidth]{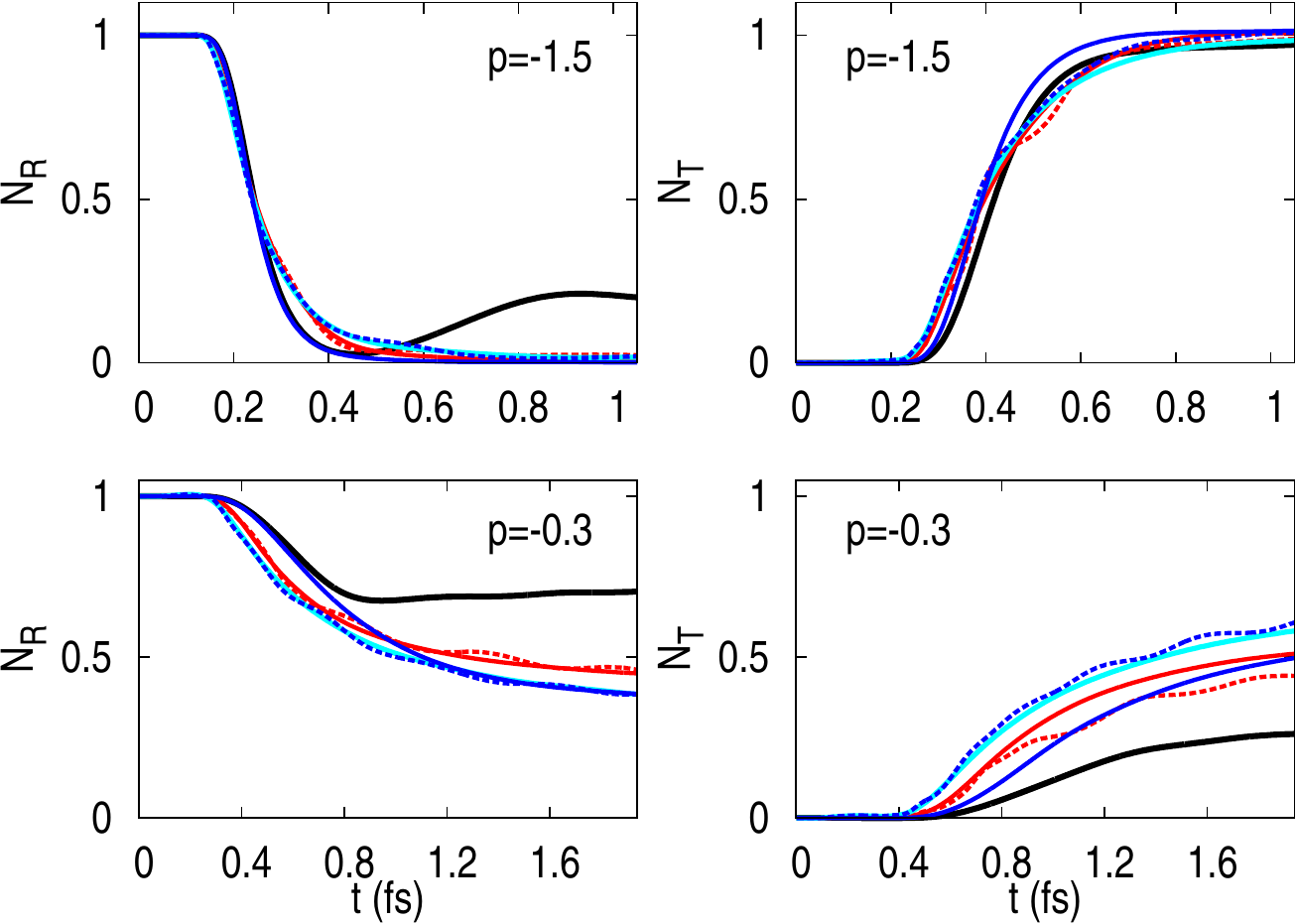}
 \caption{(color online). Number of reflected electrons $N_R$ (left panel) and transmitted electrons $N_T$ (right panel) for the exact
 (black solid), ALDA ($\Phi_0^{(1)}$: red dashed, $\Phi_0^{(2)}$: red solid), 
 $v_{\rm xc}^{\rm S}$ ($\Phi_0^{(1)}$: blue dashed, $\Phi_0^{(2)}$: blue solid),
AEXX ($\Phi_0^{(1)}$: equal to $v_{\rm xc}^{\rm S}$ (blue dashed), $\Phi_0^{(2)}$: cyan solid)
 for the two different momenta
$p=-1.5$ (upper panels) and $p=-0.3$ (lower panels). 
}
  \label{fig:Fig2}
\end{figure}

A TDDFT calculation starts with the specification of the initial KS state, for which there is considerable freedom. 
For initial ground-states, the natural choice is the KS ground-state, but for a general initial state $\Psi(0)$, one may
 pick any initial KS state $\Phi(0)$ that has the same density and first time-derivative of the density as that
of the interacting system~\cite{Leeuwen}.
Recently the impact of this choice on the xc potentials has been studied~\cite{xc3,EM12}.
We consider two natural choices of initial KS state for the scattering problem. 
The first one  is  the Slater determinant, which, for our spin-singlet state involves one doubly-occupied spatial orbital,
\ben
\Phi_0^{(1)}(x_1,x_2)
=\phi_0(x_1)\phi_0(x_2) 
\label{eqn: Phi_2}
\een
with
$\phi_0(x)=\sqrt{\frac{n_0(x)}{2}}\exp\left[i\int^x \frac{j_0(x')}{n_0(x')}dx'\right]$,
where $n_0$ and $j_0$ are the initial density and current density of the interacting system respectively. 
On the one hand, as a Slater determinant, it is in keeping with usual KS approach; on the other hand it has a 
very different form than the physical state. 
The second initial KS state we consider has the scattering form of Eq.~(\ref{eqn:Psi}), with two orbitals, 
and to reproduce $n_0$ and $\partial_t{n_0}$  of Eq.~(\ref{eqn:Psi}) we must in fact take 
\ben
\begin{split}
\Phi_0^{(2)}(x_1, x_2)=\Psi_0&(x_1, x_2)\;.
\end{split}\label{eqn: Phi_11}
\een
Each of these KS states is a valid initial state for the KS evolution, for which the
 exact xc potentials $v\xc[n;\Psi_0,\Phi_0^{(1)}](x,t)$ and $v\xc[n;\Psi_0,\Phi_0^{(2)}](x,t)$ can be found that reproduce the exact density. 
 This can be done by using the global fixed-point iteration method of Ref.~\cite{gfpim0,*gfpim1,*gfpim2}.
(For the single-spatial orbital case, it can also be obtained more simply by inverting the TDKS equation~\cite{xc1}).

\begin{figure}[h]
 \centering
 \includegraphics*[width=1.0\columnwidth]{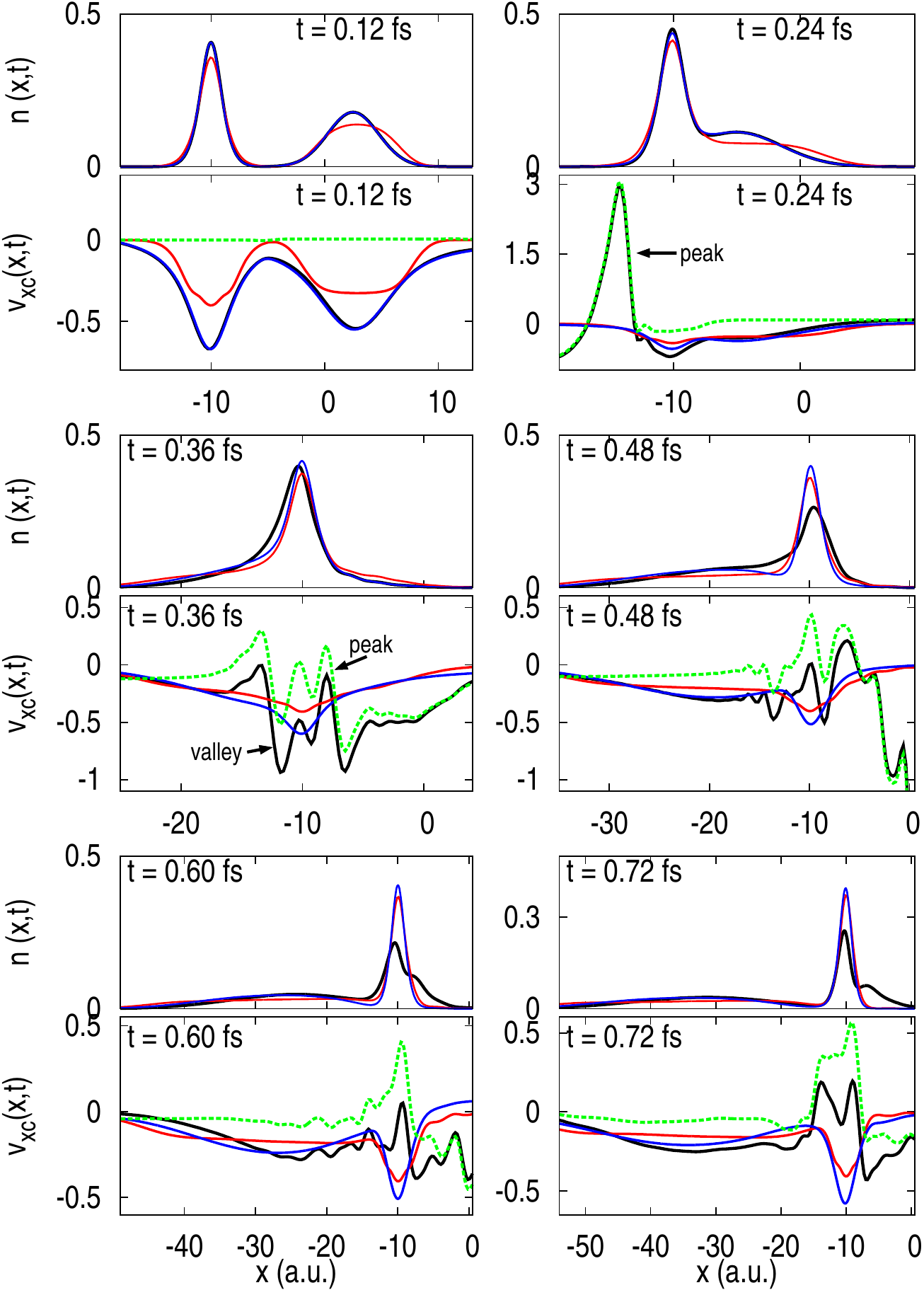}
 \caption{Same as Fig.~\ref{fig:Fig1}, but for the initial KS state $\Phi_0^{(2)}$. }
 \label{fig:Fig3}
\end{figure}

The black solid lines in the lower panels of Figs.~\ref{fig:Fig1} and~\ref{fig:Fig3} show the snapshots of the exact xc potentials
$v\xc[n;\Psi_0,\Phi_0^{(1)}](x,t)$ and $v\xc[n;\Psi_0,\Phi_0^{(2)}](x,t)$ respectively, demonstrating
 remarkable features of the xc potentials in the scattering process~\footnotemark[1].
In the  $\Phi_0^{(1)}$ case (Fig.~\ref{fig:Fig1}), $v_{\rm xc}[n;\Psi_0,\Phi_0^{(1)}] (x,t)$
develops dynamical peak and step structures throughout the scattering process, beginning from when the incident wavepacket approaches the atom's electron.
The exact xc potential for $\Phi_0^{(2)}$ (Fig.~\ref{fig:Fig3}) has no structure
 at very early times, but displays a large peak structure behind the center of the
 target during the approach (at around $t=0.24$ fs), and complicated structures after the electron reaches the interaction region. 
In fact for both choices of KS initial state, a series of peaks and valleys is evident in the interaction region, 
that are largely responsible for the scattering, as we will now argue. 

We consider propagation under two approximations in which these structures are absent. The first is the adiabatic
 local density approximation (ALDA)~\cite{1dlda}, whose density and xc potential are shown as the red lines in
 Figs.~\ref{fig:Fig1} and \ref{fig:Fig3} for the two choices of the initial state~\footnotemark[1]; see also Fig.~\ref{fig:Fig2} for 
the $N_R$ and $N_T$ values. In both cases, the ALDA dynamics fails to even qualitatively capture the reflection; for $\Phi_0^{(1)}$,
 the electron density develops unphysical spurious oscillating structures in the interaction region,
while for $\Phi_0^{(2)}$, although these oscillating structures are
 absent, and the earlier dynamics is better reproduced than the $\Phi_0^{(1)}$ case (until about $t=0.36$ fs), the ALDA xc potential too smoothly
 tracks the instantaneous density throughout. In both cases, ALDA, with its simple local dependence on
 the instantaneous density, cannot develop the peaks and valleys of the exact xc potential, and
 yields far reduced reflected density and reduced energy loss~\cite{AV06} to the target. 

One might be tempted to attribute the poor performance of ALDA for scattering simply to the 
incorrect fall-off of its xc potential, 
 however
similar results are obtained with adiabatic exact exchange AEXX (shown in Fig.~\ref{fig:Fig2}), so
 getting the
 asymptotics correct is not enough to get good dynamics~\footnotemark[1]. The next approximation arises from
 considering first the decomposition of the exact xc potential into kinetic (T) and interaction (W) terms,
 $v_{\rm xc}=v^{\rm T}_{\rm c}+v^{\rm W}_{\rm xc}$~\cite{kine1,*kine2,xc2,xc3}, where 
\ben
\begin{split}
v^{\rm T}_{\rm c}(x,t)=&\int^x \frac{1}{4n(x'',t)}\left(\frac{d}{dx'}-\frac{d}{dx''}\right)\left(\frac{d^2}{dx''^2}-\frac{d^2}{dx'^2}\right)\\
&\left(\rho_1(x',x'',t)-\rho_{1,s}(x',x'',t)\right)\vert_{x'=x''}dx'',
\end{split}\label{eqn: vcT}
\een
\ben
\begin{split}
v^{\rm W}_{\rm xc}(x,t)=\int^x dx''\int n_{\rm xc}(x',x'',t)\frac{\partial}{\partial x''}W_{ee}(|x'-x''|)dx',
\end{split}\label{eqn: vxcW}
\een
where
$n_{\rm xc}$ is the xc hole, defined via the pair density
 as $P(x',x,t)=N(N-1)\sum_{\sigma_1\cdots\sigma_N}\int\left|\Psi(x'\sigma_1,x\sigma_2,x_3\sigma_3\cdots x_N\sigma_N,t)\right|^2dx_3\cdots dx_N=
n(x,t)\left(n(x',t)+n_{\rm xc}(x',x,t)\right)$, 
and $\rho_1$ and $\rho_{1,s}$ are the spin-summed one-body density matrices
for the interacting system and KS system respectively.
For a wide range of dynamics of two-electron systems (driven dynamics
 to local excitations, charge-transfer dynamics, field-free dynamics of non-stationary states),   
$v^{\rm T}_{\rm c}$ has been found to exhibit larger non-adiabatic features than $v^{\rm W}_{\rm xc}$ and was the main origin of
 step and peak structures~\cite{xc2,xc3}. We find here that this is also true for scattering, where 
 $v^{\rm T}_{\rm c}$ largely comprises the complicated peak and valley structures in the exact
 xc potential; this is plotted as the green dotted line in Figs. \ref{fig:Fig1} and \ref{fig:Fig3}  
 (Note that the choice of  $\Phi_0^{(2)}$ makes $v_c^{\rm T}$ zero at very early times). 

The exact decomposition Eq.~(\ref{eqn: vcT})--(\ref{eqn: vxcW}) motivates 
the $v^{\rm S}_{\rm xc}$ approximation to the xc potential~\cite{xc3}, defined as
\ben
\begin{split}
v^{\rm S}_{\rm xc}(x,t)=\int^x dx''\int n^{\rm S}_{\rm xc}(x',x'',t)\frac{\partial}{\partial x''}W_{ee}(|x'-x''|)dx'\;,
\end{split}\label{eqn: vxcS}
\een
where $n^{\rm S}_{\rm xc}(x',x'',t)$ is the xc hole of the KS system. 
That is, $v^{\rm S}_{\rm xc}(x,t)$ replaces all quantities on the right of Eq.~(\ref{eqn: vcT})--(\ref{eqn: vxcW})  with their single-particle KS values. 
This approximation yields an orbital-dependent functional, that generally has spatial- and
 time- non-local dependence on the density, and 
it reduces to the time-dependent (TD) EXX approximation when the KS state is a
 Slater determinant. So, propagating $\Phi_0^{(1)}$ with $v^{\rm S}_{\rm xc}$ is equivalent to 
TDEXX, and in fact for $\Phi_0^{(1)}$, 
with two electrons in the same spatial orbital, TDEXX reduces to AEXX.
 Although this is self-interaction free and has correct asymptotic behavior, this also displays spurious
  oscillation in the density from early on as was seen in ALDA, and does not improve the
 time-resolved reflection probabilities (blue line in Figs. \ref{fig:Fig1} and~\ref{fig:Fig2})~\footnotemark[1]. 
   With the choice of  $\Phi_0^{(2)}$, $v^{\rm S}_{\rm xc}$ (which is then non-adiabatic and includes some correlation),
 reproduces the exact xc potential and dynamics very well at earlier times, as is evident in
 Fig. \ref{fig:Fig3}, and the approach of the electron to the target is very well-captured. Actually $v^{\rm S}_{\rm xc}$ 
is a very good approximation to $v^{\rm W}_{\rm xc}$, however without $v^{\rm T}_{\rm c}$, the dynamics 
begins to deviate from the exact during the scattering process, and ultimately it also gives reduced
 reflection and reduced energy loss to the target~\footnotemark[1]. 
 In fact, propagation with the exact $v_{\rm xc}^{\rm W}  = v_{\rm xc} - v^{\rm T}_{\rm c}$, for either 
initial state $\Phi_0^{(1)}$ or $\Phi_0^{(2)}$, gives reduced reflection and energy loss (Data not 
shown here. For $\Phi_0^{(2)}$, propagation with $v_{\rm xc}^{\rm W}$ yields results very similar to that with $v^{\rm S}_{\rm xc}(x,t)$.).
Thus  $v^{\rm T}_{\rm c}$ is the key component to reproduce scattering correctly, and
 approximations which lack a good model for this term will fail to even qualitatively capture the reflection. 

\begin{figure}[h]
 \centering
\includegraphics*[width=1.0\columnwidth]{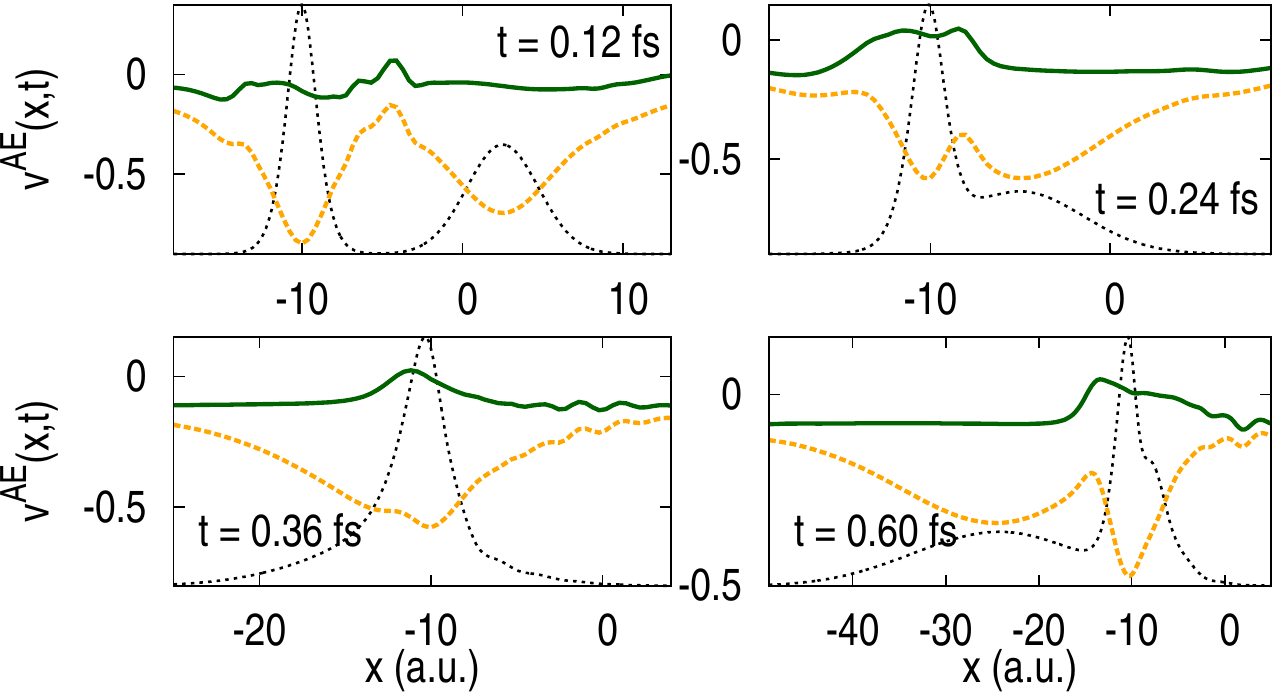}
 \caption{(color online). 
Snapshots of  $v^{\rm AE}_{\rm xc}$ (orange dotted) and $v^{\rm AE, T}_{\rm c}$ (dark-green).
Exact electron density is also shown as black dotted line.
}
  \label{fig:Fig4}
\end{figure}

Can the relevant structures in $v^{\rm T}_{\rm c}$ be modeled by any adiabatic approximation? To answer this, we consider
 the best adiabatic approximation possible, the adiabatically exact (AE). 
This is defined as the exact ground-state xc potential evaluated on the instantaneous density, 
i.e., $v^{\rm AE}_{\rm xc}[n]=v^{\rm ex. gs.}_{s}[n]-v^{\rm ex. gs.}_{\rm ext}[n]-v_{\rm H}[n]$,
where $v^{\rm ex. gs.}_{\rm ext}[n]$ ($v^{\rm ex. gs.}_{s}[n]$) is the external (KS) potential for interacting (non-interacting) electrons 
whose ground state has density $n$.
By inversion of the ground-state KS equation, $v^{\rm ex. gs.}_s[n] = \nabla^2\sqrt{n}/({2\sqrt{n}})$
 up to a constant, while to find $v^{\rm ex. gs.}_{\rm ext}[n]$, an iterative method~\cite{thiele} was employed.
In Fig.~\ref{fig:Fig4}, we show snapshots of $v^{\rm AE}_{\rm xc}$ along with the AE kinetic
 contribution $v^{\rm AE, T}_{\rm c}[n]=v^{\rm AE}_{\rm xc}[n]-v^{\rm AE, W}_{\rm xc}[n]$. 
We see that $v^{\rm AE, T}_{\rm c}$, although not structure-less, misses the dominant structures of the
 exact kinetic contribution $v^{\rm T}_{\rm c}$; these are truly non-adiabatic features and are essential to capture the scattering even qualitatively.

We consider now the case of elastic scattering by reducing the incoming momentum to $p = -0.3$ a.u, 
such that the energy is lower than the lowest excitation of the target (which is about $\omega = 0.4$ a.u.), 
so that inelastic channels are closed. Again neither ALDA nor $v\xc^{\rm S}$ even qualitatively
 capture the scattering dynamics, for either choice of initial state, as is clear from the lower panels in 
Fig.~\ref{fig:Fig2} (a movie is given in the supplementary information~\footnotemark[1]). 
Yet, Refs.~\cite{scat1,*scat2} show that good scattering cross-sections can 
be extracted from the TDDFT linear response formalism, when using standard approximations. 
 With a formally exact theory, the time-domain picture should agree 
with the time-independent picture for elastic scattering in the long-time limit~\cite{Taylorbook}, however this is not the case for approximate TDDFT.
The time-resolved picture presents scattering as a fully non-equilibrium problem, 
 where the
 system starts far from a ground-state and so from the very start the
 xc functional must be evaluated on systems whose underlying
 wavefunctions are far from ground-states. This is quite unlike using
 the same functional in a linear response calculation where it is
 evaluated on densities close to a ground-state; typical adiabatic
 approximations work much better in the latter.
As the present work has shown, non-adiabaticity beyond the
 adiabatic approximation and beyond what is contained in $v_{\rm xc}^{\rm S}$ is required to give even 
qualitatively accurate time-resolved dynamics. 
The situation is similar to that for field-free dynamics of a system in a superposition state:
one can find adiabatic functionals which yield good predictions for excitation
 energies and so can accurately predict the period of its density-oscillations, but when the time-resolved
 dynamics is run from a superposition state, the oscillation period of the time-dependent dipole can deviate significantly~\cite{xc2,FLSM15}.

In summary,
we have analyzed the exact xc potentials for a two-electron model system for both inelastic and elastic 
scattering processes. The choice of initial KS state has a significant effect on the ensuing dynamics when 
approximate functionals are used; we show that choosing a Slater determinant for the KS system results
 in spurious density oscillations, while a KS state with the same configuration as the true gives more physical results.
We revealed how and why ALDA and EXX cannot reproduce the correct scattering dynamics, and showed that although the
 recently proposed non-adiabatic approximation $v^{\rm S}_{\rm xc}$ greatly improves the dynamics up to the time of interaction,
 ultimately it also fails to capture the scattering accurately. The peak and valley structures in the
 kinetic component of the exact xc potential $v^{\rm T}_{\rm c}$ are missing in all these approximations, and 
this work suggests the urgent need for reasonable density- or orbital-functional approximations to this term (Eq.~(\ref{eqn: vcT})) to improve the reliability of
 TDDFT to describe time-resolved scattering processes. Similar trends hold for a model electron-Helium+ system (data not shown here)
, and how the effects described here scale to larger systems will be investigated in future work.

\begin{acknowledgments}
YS is supported by JSPS KAKENHI Grant No. JP16K17768.
KW is supported by JSPS KAKENHI Grant No. JP16K05483.
Financial support from the US National Science Foundation
CHE-1566197 (NTM) and the Department of Energy, Office
of Basic Energy Sciences, Division of Chemical Sciences,
Geosciences and Biosciences under Award DE-SC0015344
are also gratefully acknowledged.
Part of the computations were performed on
the supercomputers of the Institute for Solid State Physics,
The University of Tokyo.
\end{acknowledgments}

\bibliography{./scattering}

%merlin.mbs apsrev4-1.bst 2010-07-25 4.21a (PWD, AO, DPC) hacked
%Control: key (0)
%Control: author (8) initials jnrlst
%Control: editor formatted (1) identically to author
%Control: production of article title (-1) disabled
%Control: page (0) single
%Control: year (1) truncated
%Control: production of eprint (0) enabled
\begin{thebibliography}{46}%
\makeatletter
\providecommand \@ifxundefined [1]{%
 \@ifx{#1\undefined}
}%
\providecommand \@ifnum [1]{%
 \ifnum #1\expandafter \@firstoftwo
 \else \expandafter \@secondoftwo
 \fi
}%
\providecommand \@ifx [1]{%
 \ifx #1\expandafter \@firstoftwo
 \else \expandafter \@secondoftwo
 \fi
}%
\providecommand \natexlab [1]{#1}%
\providecommand \enquote  [1]{``#1''}%
\providecommand \bibnamefont  [1]{#1}%
\providecommand \bibfnamefont [1]{#1}%
\providecommand \citenamefont [1]{#1}%
\providecommand \href@noop [0]{\@secondoftwo}%
\providecommand \href [0]{\begingroup \@sanitize@url \@href}%
\providecommand \@href[1]{\@@startlink{#1}\@@href}%
\providecommand \@@href[1]{\endgroup#1\@@endlink}%
\providecommand \@sanitize@url [0]{\catcode `\\12\catcode `\$12\catcode
  `\&12\catcode `\#12\catcode `\^12\catcode `\_12\catcode `\%12\relax}%
\providecommand \@@startlink[1]{}%
\providecommand \@@endlink[0]{}%
\providecommand \url  [0]{\begingroup\@sanitize@url \@url }%
\providecommand \@url [1]{\endgroup\@href {#1}{\urlprefix }}%
\providecommand \urlprefix  [0]{URL }%
\providecommand \Eprint [0]{\href }%
\providecommand \doibase [0]{http://dx.doi.org/}%
\providecommand \selectlanguage [0]{\@gobble}%
\providecommand \bibinfo  [0]{\@secondoftwo}%
\providecommand \bibfield  [0]{\@secondoftwo}%
\providecommand \translation [1]{[#1]}%
\providecommand \BibitemOpen [0]{}%
\providecommand \bibitemStop [0]{}%
\providecommand \bibitemNoStop [0]{.\EOS\space}%
\providecommand \EOS [0]{\spacefactor3000\relax}%
\providecommand \BibitemShut  [1]{\csname bibitem#1\endcsname}%
\let\auto@bib@innerbib\@empty
%</preamble>
\bibitem [{\citenamefont {Cuevas}\ and\ \citenamefont {Scheer}(2010)}]{ME}%
  \BibitemOpen
  \bibfield  {author} {\bibinfo {author} {\bibfnamefont {J.~C.}\ \bibnamefont
  {Cuevas}}\ and\ \bibinfo {author} {\bibfnamefont {E.}~\bibnamefont
  {Scheer}},\ }\href@noop {} {\emph {\bibinfo {title} {Molecular Electronics:
  An Introduction to Theory and Experiment}}}\ (\bibinfo  {publisher} {World
  Scientific Publishing},\ \bibinfo {year} {2010})\BibitemShut {NoStop}%
\bibitem [{\citenamefont {Krausz}\ and\ \citenamefont {Ivanov}(2009)}]{SFP}%
  \BibitemOpen
  \bibfield  {author} {\bibinfo {author} {\bibfnamefont {F.}~\bibnamefont
  {Krausz}}\ and\ \bibinfo {author} {\bibfnamefont {M.}~\bibnamefont
  {Ivanov}},\ }\href@noop {} {\bibfield  {journal} {\bibinfo  {journal} {Rev.
  Mod. Phys.}\ }\textbf {\bibinfo {volume} {81}},\ \bibinfo {pages} {163}
  (\bibinfo {year} {2009})}\BibitemShut {NoStop}%
\bibitem [{\citenamefont {Meyer}\ \emph {et~al.}(2008)\citenamefont {Meyer},
  \citenamefont {Girit}, \citenamefont {Crommie},\ and\ \citenamefont
  {Zettl}}]{TEM}%
  \BibitemOpen
  \bibfield  {author} {\bibinfo {author} {\bibfnamefont {J.~C.}\ \bibnamefont
  {Meyer}}, \bibinfo {author} {\bibfnamefont {C.~O.}\ \bibnamefont {Girit}},
  \bibinfo {author} {\bibfnamefont {M.~F.}\ \bibnamefont {Crommie}}, \ and\
  \bibinfo {author} {\bibfnamefont {A.}~\bibnamefont {Zettl}},\ }\href@noop {}
  {\bibfield  {journal} {\bibinfo  {journal} {Nature}\ }\textbf {\bibinfo
  {volume} {454}},\ \bibinfo {pages} {319} (\bibinfo {year}
  {2008})}\BibitemShut {NoStop}%
\bibitem [{\citenamefont {Ciston}\ \emph {et~al.}(2015)\citenamefont {Ciston},
  \citenamefont {Brown}, \citenamefont {D'Alfonso}, \citenamefont {Koirala},
  \citenamefont {Ophus}, \citenamefont {Lin}, \citenamefont {Suzuki},
  \citenamefont {Inada}, \citenamefont {Zhu}, \citenamefont {Allen},\ and\
  \citenamefont {Marks}}]{SEM}%
  \BibitemOpen
  \bibfield  {author} {\bibinfo {author} {\bibfnamefont {J.}~\bibnamefont
  {Ciston}}, \bibinfo {author} {\bibfnamefont {H.~G.}\ \bibnamefont {Brown}},
  \bibinfo {author} {\bibfnamefont {A.~J.}\ \bibnamefont {D'Alfonso}}, \bibinfo
  {author} {\bibfnamefont {P.}~\bibnamefont {Koirala}}, \bibinfo {author}
  {\bibfnamefont {C.}~\bibnamefont {Ophus}}, \bibinfo {author} {\bibfnamefont
  {Y.}~\bibnamefont {Lin}}, \bibinfo {author} {\bibfnamefont {Y.}~\bibnamefont
  {Suzuki}}, \bibinfo {author} {\bibfnamefont {H.}~\bibnamefont {Inada}},
  \bibinfo {author} {\bibfnamefont {Y.}~\bibnamefont {Zhu}}, \bibinfo {author}
  {\bibfnamefont {L.~J.}\ \bibnamefont {Allen}}, \ and\ \bibinfo {author}
  {\bibfnamefont {L.~D.}\ \bibnamefont {Marks}},\ }\href@noop {} {\bibfield
  {journal} {\bibinfo  {journal} {Nat. Commn,}\ }\textbf {\bibinfo {volume}
  {6}},\ \bibinfo {pages} {7358} (\bibinfo {year} {2015})}\BibitemShut
  {NoStop}%
\bibitem [{\citenamefont {Bouda{\"i}ffa}\ \emph {et~al.}(2000)\citenamefont
  {Bouda{\"i}ffa}, \citenamefont {Cloutier}, \citenamefont {Hunting},
  \citenamefont {Huels},\ and\ \citenamefont {Sanche}}]{DNA}%
  \BibitemOpen
  \bibfield  {author} {\bibinfo {author} {\bibfnamefont {B.}~\bibnamefont
  {Bouda{\"i}ffa}}, \bibinfo {author} {\bibfnamefont {P.}~\bibnamefont
  {Cloutier}}, \bibinfo {author} {\bibfnamefont {D.}~\bibnamefont {Hunting}},
  \bibinfo {author} {\bibfnamefont {M.~A.}\ \bibnamefont {Huels}}, \ and\
  \bibinfo {author} {\bibfnamefont {L.}~\bibnamefont {Sanche}},\ }\href@noop {}
  {\bibfield  {journal} {\bibinfo  {journal} {Science}\ }\textbf {\bibinfo
  {volume} {287}},\ \bibinfo {pages} {1658} (\bibinfo {year}
  {2000})}\BibitemShut {NoStop}%
\bibitem [{\citenamefont {Runge}\ and\ \citenamefont {Gross}(1984)}]{tddft1}%
  \BibitemOpen
  \bibfield  {author} {\bibinfo {author} {\bibfnamefont {E.}~\bibnamefont
  {Runge}}\ and\ \bibinfo {author} {\bibfnamefont {E.~K.~U.}\ \bibnamefont
  {Gross}},\ }\href@noop {} {\bibfield  {journal} {\bibinfo  {journal} {Phys.
  Rev. Lett.}\ }\textbf {\bibinfo {volume} {52}},\ \bibinfo {pages} {997}
  (\bibinfo {year} {1984})}\BibitemShut {NoStop}%
\bibitem [{\citenamefont {Ullrich}(2012)}]{tddft2}%
  \BibitemOpen
  \bibfield  {author} {\bibinfo {author} {\bibfnamefont {C.~A.}\ \bibnamefont
  {Ullrich}},\ }\href@noop {} {\emph {\bibinfo {title} {Time-Dependent
  Density-Functional Theory: Concepts and Applications}}}\ (\bibinfo
  {publisher} {Oxford University Press},\ \bibinfo {year} {2012})\BibitemShut
  {NoStop}%
\bibitem [{\citenamefont {Maitra}(2016)}]{tddft3}%
  \BibitemOpen
  \bibfield  {author} {\bibinfo {author} {\bibfnamefont {N.~T.}\ \bibnamefont
  {Maitra}},\ }\href@noop {} {\bibfield  {journal} {\bibinfo  {journal} {J.
  Chem. Phys.}\ }\textbf {\bibinfo {volume} {144}},\ \bibinfo {pages} {220901}
  (\bibinfo {year} {2016})}\BibitemShut {NoStop}%
\bibitem [{\citenamefont {Rozzi}\ \emph {et~al.}(2013)\citenamefont {Rozzi},
  \citenamefont {Falke}, \citenamefont {Spallanzani}, \citenamefont {Rubio},
  \citenamefont {Molinari}, \citenamefont {Brida}, \citenamefont {Maiuri},
  \citenamefont {Cerullo}, \citenamefont {Schramm}, \citenamefont
  {Christoffers},\ and\ \citenamefont {Lienau}}]{tddft4}%
  \BibitemOpen
  \bibfield  {author} {\bibinfo {author} {\bibfnamefont {C.~A.}\ \bibnamefont
  {Rozzi}}, \bibinfo {author} {\bibfnamefont {S.~M.}\ \bibnamefont {Falke}},
  \bibinfo {author} {\bibfnamefont {N.}~\bibnamefont {Spallanzani}}, \bibinfo
  {author} {\bibfnamefont {A.}~\bibnamefont {Rubio}}, \bibinfo {author}
  {\bibfnamefont {E.}~\bibnamefont {Molinari}}, \bibinfo {author}
  {\bibfnamefont {D.}~\bibnamefont {Brida}}, \bibinfo {author} {\bibfnamefont
  {M.}~\bibnamefont {Maiuri}}, \bibinfo {author} {\bibfnamefont
  {G.}~\bibnamefont {Cerullo}}, \bibinfo {author} {\bibfnamefont
  {H.}~\bibnamefont {Schramm}}, \bibinfo {author} {\bibfnamefont
  {J.}~\bibnamefont {Christoffers}}, \ and\ \bibinfo {author} {\bibfnamefont
  {C.}~\bibnamefont {Lienau}},\ }\href@noop {} {\bibfield  {journal} {\bibinfo
  {journal} {Nat. Commun.}\ }\textbf {\bibinfo {volume} {4}},\ \bibinfo {pages}
  {1602} (\bibinfo {year} {2013})}\BibitemShut {NoStop}%
\bibitem [{\citenamefont {Penka Fowe}\ and\ \citenamefont {Bandrauk}(2011)}]{tddft5}%
  \BibitemOpen
  \bibfield  {author} {\bibinfo {author} {\bibfnamefont {E.}\ \bibnamefont
  {Penka Fowe}}\ and\ \bibinfo {author} {\bibfnamefont {A.~D.}\ \bibnamefont
  {Bandrauk}},\ }\href@noop {} {\bibfield  {journal} {\bibinfo  {journal}
  {Phys. Rev. A}\ }\textbf {\bibinfo {volume} {84}},\ \bibinfo {pages} {035402}
  (\bibinfo {year} {2011})}\BibitemShut {NoStop}%
\bibitem [{\citenamefont {Yabana}\ \emph {et~al.}(2012)\citenamefont {Yabana},
  \citenamefont {Sugiyama}, \citenamefont {Shinohara}, \citenamefont {Otobe},\
  and\ \citenamefont {Bertsch}}]{tddft6}%
  \BibitemOpen
  \bibfield  {author} {\bibinfo {author} {\bibfnamefont {K.}~\bibnamefont
  {Yabana}}, \bibinfo {author} {\bibfnamefont {T.}~\bibnamefont {Sugiyama}},
  \bibinfo {author} {\bibfnamefont {Y.}~\bibnamefont {Shinohara}}, \bibinfo
  {author} {\bibfnamefont {T.}~\bibnamefont {Otobe}}, \ and\ \bibinfo {author}
  {\bibfnamefont {G.~F.}\ \bibnamefont {Bertsch}},\ }\href@noop {} {\bibfield
  {journal} {\bibinfo  {journal} {Phys. Rev. B}\ }\textbf {\bibinfo {volume}
  {85}},\ \bibinfo {pages} {045134} (\bibinfo {year} {2012})}\BibitemShut
  {NoStop}%
\bibitem [{\citenamefont {Castro}\ \emph {et~al.}(2012)\citenamefont {Castro},
  \citenamefont {Werschnik},\ and\ \citenamefont {Gross}}]{tddft7}%
  \BibitemOpen
  \bibfield  {author} {\bibinfo {author} {\bibfnamefont {A.}~\bibnamefont
  {Castro}}, \bibinfo {author} {\bibfnamefont {J.}~\bibnamefont {Werschnik}}, \
  and\ \bibinfo {author} {\bibfnamefont {E.~K.~U.}\ \bibnamefont {Gross}},\
  }\href@noop {} {\bibfield  {journal} {\bibinfo  {journal} {Phys. Rev. Lett.}\
  }\textbf {\bibinfo {volume} {109}},\ \bibinfo {pages} {153603} (\bibinfo
  {year} {2012})}\BibitemShut {NoStop}%
\bibitem [{\citenamefont {Miyamoto}\ \emph {et~al.}(2015)\citenamefont
  {Miyamoto}, \citenamefont {Zhang}, \citenamefont {Miyazaki},\ and\
  \citenamefont {Rubio}}]{tddft8}%
  \BibitemOpen
  \bibfield  {author} {\bibinfo {author} {\bibfnamefont {Y.}~\bibnamefont
  {Miyamoto}}, \bibinfo {author} {\bibfnamefont {H.}~\bibnamefont {Zhang}},
  \bibinfo {author} {\bibfnamefont {T.}~\bibnamefont {Miyazaki}}, \ and\
  \bibinfo {author} {\bibfnamefont {A.}~\bibnamefont {Rubio}},\ }\href@noop {}
  {\bibfield  {journal} {\bibinfo  {journal} {Phys. Rev. Lett.}\ }\textbf
  {\bibinfo {volume} {114}},\ \bibinfo {pages} {116102} (\bibinfo {year}
  {2015})}\BibitemShut {NoStop}%
\bibitem [{\citenamefont {Wang}\ \emph {et~al.}(2015)\citenamefont {Wang},
  \citenamefont {Li},\ and\ \citenamefont {Wang}}]{tddft9}%
  \BibitemOpen
  \bibfield  {author} {\bibinfo {author} {\bibfnamefont {Z.}~\bibnamefont
  {Wang}}, \bibinfo {author} {\bibfnamefont {S.-S.}\ \bibnamefont {Li}}, \ and\
  \bibinfo {author} {\bibfnamefont {L.-W.}\ \bibnamefont {Wang}},\ }\href@noop
  {} {\bibfield  {journal} {\bibinfo  {journal} {Phys. Rev. Lett.}\ }\textbf
  {\bibinfo {volume} {114}},\ \bibinfo {pages} {063004} (\bibinfo {year}
  {2015})}\BibitemShut {NoStop}%
\bibitem [{\citenamefont {Elliott}\ \emph {et~al.}(2016)\citenamefont
  {Elliott}, \citenamefont {Krieger}, \citenamefont {Dewhurst}, \citenamefont
  {Sharma},\ and\ \citenamefont {Gross}}]{tddft10}%
  \BibitemOpen
  \bibfield  {author} {\bibinfo {author} {\bibfnamefont {P.}~\bibnamefont
  {Elliott}}, \bibinfo {author} {\bibfnamefont {K.}~\bibnamefont {Krieger}},
  \bibinfo {author} {\bibfnamefont {J.~K.}\ \bibnamefont {Dewhurst}}, \bibinfo
  {author} {\bibfnamefont {S.}~\bibnamefont {Sharma}}, \ and\ \bibinfo {author}
  {\bibfnamefont {E.~K.~U.}\ \bibnamefont {Gross}},\ }\href@noop {} {\bibfield
  {journal} {\bibinfo  {journal} {New J. Phys.}\ }\textbf {\bibinfo {volume}
  {18}},\ \bibinfo {pages} {013014} (\bibinfo {year} {2016})}\BibitemShut
  {NoStop}%
\bibitem [{\citenamefont {van Faassen}\ \emph {et~al.}(2007)\citenamefont {van
  Faassen}, \citenamefont {Wasserman}, \citenamefont {Engel}, \citenamefont
  {Zhang},\ and\ \citenamefont {Burke}}]{scat1}%
  \BibitemOpen
  \bibfield  {author} {\bibinfo {author} {\bibfnamefont {M.}~\bibnamefont {van
  Faassen}}, \bibinfo {author} {\bibfnamefont {A.}~\bibnamefont {Wasserman}},
  \bibinfo {author} {\bibfnamefont {E.}~\bibnamefont {Engel}}, \bibinfo
  {author} {\bibfnamefont {F.}~\bibnamefont {Zhang}}, \ and\ \bibinfo {author}
  {\bibfnamefont {K.}~\bibnamefont {Burke}},\ }\href@noop {} {\bibfield
  {journal} {\bibinfo  {journal} {Phys. Rev. Lett.}\ }\textbf {\bibinfo
  {volume} {99}},\ \bibinfo {pages} {043005} (\bibinfo {year}
  {2007})}\BibitemShut {NoStop}%
\bibitem [{\citenamefont {van Faassen}\ and\ \citenamefont
  {Burke}(2009)}]{scat2}%
  \BibitemOpen
  \bibfield  {author} {\bibinfo {author} {\bibfnamefont {M.}~\bibnamefont {van
  Faassen}}\ and\ \bibinfo {author} {\bibfnamefont {K.}~\bibnamefont {Burke}},\
  }\href@noop {} {\bibfield  {journal} {\bibinfo  {journal} {Phys. Chem. Chem.
  Phys.}\ }\textbf {\bibinfo {volume} {11}},\ \bibinfo {pages} {4437} (\bibinfo
  {year} {2009})}\BibitemShut {NoStop}%
\bibitem [{\citenamefont {Wasserman}\ \emph {et~al.}(2005)\citenamefont
  {Wasserman}, \citenamefont {Maitra},\ and\ \citenamefont {Burke}}]{WMB05}%
  \BibitemOpen
  \bibfield  {author} {\bibinfo {author} {\bibfnamefont {A.}~\bibnamefont
  {Wasserman}}, \bibinfo {author} {\bibfnamefont {N.~T.}\ \bibnamefont
  {Maitra}}, \ and\ \bibinfo {author} {\bibfnamefont {K.}~\bibnamefont
  {Burke}},\ }\href@noop {} {\bibfield  {journal} {\bibinfo  {journal} {J.
  Chem. Phys.}\ }\textbf {\bibinfo {volume} {122}},\ \bibinfo {pages} {144103}
  (\bibinfo {year} {2005})}\BibitemShut {NoStop}%
\bibitem [{\citenamefont {Gao}\ \emph {et~al.}(2014)\citenamefont {Gao},
  \citenamefont {Wang}, \citenamefont {Wang},\ and\ \citenamefont
  {Zhang}}]{GWWZ14}%
  \BibitemOpen
  \bibfield  {author} {\bibinfo {author} {\bibfnamefont {C.-Z.}\ \bibnamefont
  {Gao}}, \bibinfo {author} {\bibfnamefont {J.}~\bibnamefont {Wang}}, \bibinfo
  {author} {\bibfnamefont {F.}~\bibnamefont {Wang}}, \ and\ \bibinfo {author}
  {\bibfnamefont {F.-S.}\ \bibnamefont {Zhang}},\ }\href@noop {} {\bibfield
  {journal} {\bibinfo  {journal} {J. Chem. Phys.}\ }\textbf {\bibinfo {volume}
  {140}},\ \bibinfo {pages} {054308} (\bibinfo {year} {2014})}\BibitemShut
  {NoStop}%
\bibitem [{\citenamefont {Quashie}\ \emph {et~al.}(2017)\citenamefont
  {Quashie}, \citenamefont {Saha}, \citenamefont {Andrade},\ and\ \citenamefont
  {Correa}}]{QSAC17}%
  \BibitemOpen
  \bibfield  {author} {\bibinfo {author} {\bibfnamefont {E.~E.}\ \bibnamefont
  {Quashie}}, \bibinfo {author} {\bibfnamefont {B.~C.}\ \bibnamefont {Saha}},
  \bibinfo {author} {\bibfnamefont {X.}~\bibnamefont {Andrade}}, \ and\
  \bibinfo {author} {\bibfnamefont {A.~A.}\ \bibnamefont {Correa}},\
  }\href@noop {} {\bibfield  {journal} {\bibinfo  {journal} {Phys. Rev. A}\
  }\textbf {\bibinfo {volume} {95}},\ \bibinfo {pages} {042517} (\bibinfo
  {year} {2017})}\BibitemShut {NoStop}%
\bibitem [{\citenamefont {Tsubonoya}\ \emph {et~al.}(2014)\citenamefont
  {Tsubonoya}, \citenamefont {Hu},\ and\ \citenamefont {Watanabe}}]{scat3}%
  \BibitemOpen
  \bibfield  {author} {\bibinfo {author} {\bibfnamefont {K.}~\bibnamefont
  {Tsubonoya}}, \bibinfo {author} {\bibfnamefont {C.}~\bibnamefont {Hu}}, \
  and\ \bibinfo {author} {\bibfnamefont {K.}~\bibnamefont {Watanabe}},\
  }\href@noop {} {\bibfield  {journal} {\bibinfo  {journal} {Phys. Rev. B}\
  }\textbf {\bibinfo {volume} {90}},\ \bibinfo {pages} {035416} (\bibinfo
  {year} {2014})}\BibitemShut {NoStop}%
\bibitem [{\citenamefont {Ueda}\ \emph {et~al.}(2016)\citenamefont {Ueda},
  \citenamefont {Suzuki},\ and\ \citenamefont {Watanabe}}]{scat4}%
  \BibitemOpen
  \bibfield  {author} {\bibinfo {author} {\bibfnamefont {Y.}~\bibnamefont
  {Ueda}}, \bibinfo {author} {\bibfnamefont {Y.}~\bibnamefont {Suzuki}}, \ and\
  \bibinfo {author} {\bibfnamefont {K.}~\bibnamefont {Watanabe}},\ }\href@noop
  {} {\bibfield  {journal} {\bibinfo  {journal} {Phys. Rev. B}\ }\textbf
  {\bibinfo {volume} {94}},\ \bibinfo {pages} {035403} (\bibinfo {year}
  {2016})}\BibitemShut {NoStop}%
\bibitem [{\citenamefont {Miyauchi}\ \emph {et~al.}(2017)\citenamefont
  {Miyauchi}, \citenamefont {Ueda}, \citenamefont {Suzuki},\ and\ \citenamefont
  {Watanabe}}]{scat5}%
  \BibitemOpen
  \bibfield  {author} {\bibinfo {author} {\bibfnamefont {H.}~\bibnamefont
  {Miyauchi}}, \bibinfo {author} {\bibfnamefont {Y.}~\bibnamefont {Ueda}},
  \bibinfo {author} {\bibfnamefont {Y.}~\bibnamefont {Suzuki}}, \ and\ \bibinfo
  {author} {\bibfnamefont {K.}~\bibnamefont {Watanabe}},\ }\href@noop {}
  {\bibfield  {journal} {\bibinfo  {journal} {Phys. Rev. B}\ }\textbf {\bibinfo
  {volume} {95}},\ \bibinfo {pages} {125425} (\bibinfo {year}
  {2017})}\BibitemShut {NoStop}%
\bibitem [{\citenamefont {Da}\ \emph {et~al.}(2017)\citenamefont {Da},
  \citenamefont {Liu}, \citenamefont {Yamamoto}, \citenamefont {Ueda},
  \citenamefont {Watanabe}, \citenamefont {Cuong}, \citenamefont {Li},
  \citenamefont {Tsukagoshi}, \citenamefont {Yoshikawa}, \citenamefont {Iwai},
  \citenamefont {Tanuma}, \citenamefont {Guo}, \citenamefont {Gao},
  \citenamefont {Sun},\ and\ \citenamefont {Ding}}]{scat6}%
  \BibitemOpen
  \bibfield  {author} {\bibinfo {author} {\bibfnamefont {B.}~\bibnamefont
  {Da}}, \bibinfo {author} {\bibfnamefont {J.}~\bibnamefont {Liu}}, \bibinfo
  {author} {\bibfnamefont {M.}~\bibnamefont {Yamamoto}}, \bibinfo {author}
  {\bibfnamefont {Y.}~\bibnamefont {Ueda}}, \bibinfo {author} {\bibfnamefont
  {K.}~\bibnamefont {Watanabe}}, \bibinfo {author} {\bibfnamefont {N.~T.}\
  \bibnamefont {Cuong}}, \bibinfo {author} {\bibfnamefont {S.}~\bibnamefont
  {Li}}, \bibinfo {author} {\bibfnamefont {K.}~\bibnamefont {Tsukagoshi}},
  \bibinfo {author} {\bibfnamefont {H.}~\bibnamefont {Yoshikawa}}, \bibinfo
  {author} {\bibfnamefont {H.}~\bibnamefont {Iwai}}, \bibinfo {author}
  {\bibfnamefont {S.}~\bibnamefont {Tanuma}}, \bibinfo {author} {\bibfnamefont
  {H.}~\bibnamefont {Guo}}, \bibinfo {author} {\bibfnamefont {Z.}~\bibnamefont
  {Gao}}, \bibinfo {author} {\bibfnamefont {X.}~\bibnamefont {Sun}}, \ and\
  \bibinfo {author} {\bibfnamefont {Z.}~\bibnamefont {Ding}},\ }\href@noop {}
  {\bibfield  {journal} {\bibinfo  {journal} {Nat. Commun.}\ }\textbf {\bibinfo
  {volume} {8}},\ \bibinfo {pages} {15629} (\bibinfo {year}
  {2017})}\BibitemShut {NoStop}%
\bibitem [{\citenamefont {Raghunathan}\ and\ \citenamefont
  {Nest}(2012)}]{fail1}%
  \BibitemOpen
  \bibfield  {author} {\bibinfo {author} {\bibfnamefont {S.}~\bibnamefont
  {Raghunathan}}\ and\ \bibinfo {author} {\bibfnamefont {M.}~\bibnamefont
  {Nest}},\ }\href@noop {} {\bibfield  {journal} {\bibinfo  {journal} {J. Chem.
  Theory Comput.}\ }\textbf {\bibinfo {volume} {8}},\ \bibinfo {pages} {806}
  (\bibinfo {year} {2012})}\BibitemShut {NoStop}%
\bibitem [{\citenamefont {Habenicht}\ \emph {et~al.}(2014)\citenamefont
  {Habenicht}, \citenamefont {Tani}, \citenamefont {Provorse},\ and\
  \citenamefont {Isborn}}]{fail2}%
  \BibitemOpen
  \bibfield  {author} {\bibinfo {author} {\bibfnamefont {B.~F.}\ \bibnamefont
  {Habenicht}}, \bibinfo {author} {\bibfnamefont {N.~P.}\ \bibnamefont {Tani}},
  \bibinfo {author} {\bibfnamefont {M.~R.}\ \bibnamefont {Provorse}}, \ and\
  \bibinfo {author} {\bibfnamefont {C.~M.}\ \bibnamefont {Isborn}},\
  }\href@noop {} {\bibfield  {journal} {\bibinfo  {journal} {J. Chem. Phys.}\
  }\textbf {\bibinfo {volume} {141}},\ \bibinfo {pages} {184112} (\bibinfo
  {year} {2014})}\BibitemShut {NoStop}%
\bibitem [{\citenamefont {Provorse}\ and\ \citenamefont
  {Isborn}(2016)}]{fail3}%
  \BibitemOpen
  \bibfield  {author} {\bibinfo {author} {\bibfnamefont {M.~R.}\ \bibnamefont
  {Provorse}}\ and\ \bibinfo {author} {\bibfnamefont {C.~M.}\ \bibnamefont
  {Isborn}},\ }\href@noop {} {\bibfield  {journal} {\bibinfo  {journal} {Int.
  J. Quantum. Chem.}\ }\textbf {\bibinfo {volume} {116}},\ \bibinfo {pages}
  {739} (\bibinfo {year} {2016})}\BibitemShut {NoStop}%
\bibitem [{\citenamefont {Wijewardane}\ and\ \citenamefont
  {Ullrich}(2005)}]{fail4}%
  \BibitemOpen
  \bibfield  {author} {\bibinfo {author} {\bibfnamefont {H.~O.}\ \bibnamefont
  {Wijewardane}}\ and\ \bibinfo {author} {\bibfnamefont {C.~A.}\ \bibnamefont
  {Ullrich}},\ }\href@noop {} {\bibfield  {journal} {\bibinfo  {journal} {Phys.
  Rev. Lett.}\ }\textbf {\bibinfo {volume} {95}},\ \bibinfo {pages} {086401}
  (\bibinfo {year} {2005})}\BibitemShut {NoStop}%
\bibitem [{\citenamefont {Elliott}\ \emph {et~al.}(2012)\citenamefont
  {Elliott}, \citenamefont {Fuks}, \citenamefont {Rubio},\ and\ \citenamefont
  {Maitra}}]{xc1}%
  \BibitemOpen
  \bibfield  {author} {\bibinfo {author} {\bibfnamefont {P.}~\bibnamefont
  {Elliott}}, \bibinfo {author} {\bibfnamefont {J.~I.}\ \bibnamefont {Fuks}},
  \bibinfo {author} {\bibfnamefont {A.}~\bibnamefont {Rubio}}, \ and\ \bibinfo
  {author} {\bibfnamefont {N.~T.}\ \bibnamefont {Maitra}},\ }\href@noop {}
  {\bibfield  {journal} {\bibinfo  {journal} {Phys. Rev. Lett.}\ }\textbf
  {\bibinfo {volume} {109}},\ \bibinfo {pages} {266404} (\bibinfo {year}
  {2012})}\BibitemShut {NoStop}%
\bibitem [{\citenamefont {Luo}\ \emph {et~al.}(2014)\citenamefont {Luo},
  \citenamefont {Fuks}, \citenamefont {Sandoval}, \citenamefont {Elliott},\
  and\ \citenamefont {Maitra}}]{xc2}%
  \BibitemOpen
  \bibfield  {author} {\bibinfo {author} {\bibfnamefont {K.}~\bibnamefont
  {Luo}}, \bibinfo {author} {\bibfnamefont {J.~I.}\ \bibnamefont {Fuks}},
  \bibinfo {author} {\bibfnamefont {E.~D.}\ \bibnamefont {Sandoval}}, \bibinfo
  {author} {\bibfnamefont {P.}~\bibnamefont {Elliott}}, \ and\ \bibinfo
  {author} {\bibfnamefont {N.~T.}\ \bibnamefont {Maitra}},\ }\href@noop {}
  {\bibfield  {journal} {\bibinfo  {journal} {J. Chem. Phys.}\ }\textbf
  {\bibinfo {volume} {140}},\ \bibinfo {pages} {18A515} (\bibinfo {year}
  {2014})}\BibitemShut {NoStop}%
\bibitem [{\citenamefont {Fuks}\ \emph {et~al.}(2016)\citenamefont {Fuks},
  \citenamefont {Nielsen}, \citenamefont {Ruggenthaler},\ and\ \citenamefont
  {Maitra}}]{xc3}%
  \BibitemOpen
  \bibfield  {author} {\bibinfo {author} {\bibfnamefont {J.~I.}\ \bibnamefont
  {Fuks}}, \bibinfo {author} {\bibfnamefont {S.~E.~B.}\ \bibnamefont
  {Nielsen}}, \bibinfo {author} {\bibfnamefont {M.}~\bibnamefont
  {Ruggenthaler}}, \ and\ \bibinfo {author} {\bibfnamefont {N.~T.}\
  \bibnamefont {Maitra}},\ }\href@noop {} {\bibfield  {journal} {\bibinfo
  {journal} {Phys. Chem. Chem. Phys.}\ }\textbf {\bibinfo {volume} {18}},\
  \bibinfo {pages} {20976} (\bibinfo {year} {2016})}\BibitemShut {NoStop}%
\bibitem [{\citenamefont {Ramsden}\ and\ \citenamefont {Godby}(2012)}]{RG12}%
  \BibitemOpen
  \bibfield  {author} {\bibinfo {author} {\bibfnamefont {J.~D.}\ \bibnamefont
  {Ramsden}}\ and\ \bibinfo {author} {\bibfnamefont {R.~W.}\ \bibnamefont
  {Godby}},\ }\href@noop {} {\bibfield  {journal} {\bibinfo  {journal} {Phys.
  Rev. Lett.}\ }\textbf {\bibinfo {volume} {109}},\ \bibinfo {pages} {036402}
  (\bibinfo {year} {2012})}\BibitemShut {NoStop}%
\bibitem [{\citenamefont {Javanainen}\ \emph {et~al.}(1988)\citenamefont
  {Javanainen}, \citenamefont {Eberly},\ and\ \citenamefont {Su}}]{soft}%
  \BibitemOpen
  \bibfield  {author} {\bibinfo {author} {\bibfnamefont {J.}~\bibnamefont
  {Javanainen}}, \bibinfo {author} {\bibfnamefont {J.~H.}\ \bibnamefont
  {Eberly}}, \ and\ \bibinfo {author} {\bibfnamefont {Q.}~\bibnamefont {Su}},\
  }\href@noop {} {\bibfield  {journal} {\bibinfo  {journal} {Phys. Rev. A}\
  }\textbf {\bibinfo {volume} {38}},\ \bibinfo {pages} {3430} (\bibinfo {year}
  {1988})}\BibitemShut {NoStop}%
\bibitem [{Note1()}]{Note1}%
  \BibitemOpen
  \bibinfo {note} {Movies of the dynamics are given in the Supplemental
  Material.}\BibitemShut {Stop}%
\bibitem [{\citenamefont {van Leeuwen}(1999)}]{Leeuwen}%
  \BibitemOpen
  \bibfield  {author} {\bibinfo {author} {\bibfnamefont {R.}~\bibnamefont {van
  Leeuwen}},\ }\href@noop {} {\bibfield  {journal} {\bibinfo  {journal} {Phys.
  Rev. Lett.}\ }\textbf {\bibinfo {volume} {82}},\ \bibinfo {pages} {3863}
  (\bibinfo {year} {1999})}\BibitemShut {NoStop}%
\bibitem [{\citenamefont {Elliott}\ and\ \citenamefont {Maitra}(2012)}]{EM12}%
  \BibitemOpen
  \bibfield  {author} {\bibinfo {author} {\bibfnamefont {P.}~\bibnamefont
  {Elliott}}\ and\ \bibinfo {author} {\bibfnamefont {N.~T.}\ \bibnamefont
  {Maitra}},\ }\href@noop {} {\bibfield  {journal} {\bibinfo  {journal} {Phys.
  Rev. A}\ }\textbf {\bibinfo {volume} {85}},\ \bibinfo {pages} {052510}
  (\bibinfo {year} {2012})}\BibitemShut {NoStop}%
\bibitem [{\citenamefont {Ruggenthaler}\ and\ \citenamefont {van
  Leeuwen}(2011)}]{gfpim0}%
  \BibitemOpen
  \bibfield  {author} {\bibinfo {author} {\bibfnamefont {M.}~\bibnamefont
  {Ruggenthaler}}\ and\ \bibinfo {author} {\bibfnamefont {R.}~\bibnamefont {van
  Leeuwen}},\ }\href@noop {} {\bibfield  {journal} {\bibinfo  {journal}
  {Europhys. Lett.}\ }\textbf {\bibinfo {volume} {95}},\ \bibinfo {pages}
  {13001} (\bibinfo {year} {2011})}\BibitemShut {NoStop}%
\bibitem [{\citenamefont {Ruggenthaler}\ \emph {et~al.}(2015)\citenamefont
  {Ruggenthaler}, \citenamefont {Penz},\ and\ \citenamefont {van
  Leeuwen}}]{gfpim1}%
  \BibitemOpen
  \bibfield  {author} {\bibinfo {author} {\bibfnamefont {M.}~\bibnamefont
  {Ruggenthaler}}, \bibinfo {author} {\bibfnamefont {M.}~\bibnamefont {Penz}},
  \ and\ \bibinfo {author} {\bibfnamefont {R.}~\bibnamefont {van Leeuwen}},\
  }\href@noop {} {\bibfield  {journal} {\bibinfo  {journal} {J. Phys.: Condens.
  Matter}\ }\textbf {\bibinfo {volume} {27}},\ \bibinfo {pages} {203202}
  (\bibinfo {year} {2015})}\BibitemShut {NoStop}%
\bibitem [{\citenamefont {Nielsen}\ \emph {et~al.}(2013)\citenamefont
  {Nielsen}, \citenamefont {Ruggenthaler},\ and\ \citenamefont {van
  Leeuwen}}]{gfpim2}%
  \BibitemOpen
  \bibfield  {author} {\bibinfo {author} {\bibfnamefont {S.~E.~B.}\
  \bibnamefont {Nielsen}}, \bibinfo {author} {\bibfnamefont {M.}~\bibnamefont
  {Ruggenthaler}}, \ and\ \bibinfo {author} {\bibfnamefont {R.}~\bibnamefont
  {van Leeuwen}},\ }\href@noop {} {\bibfield  {journal} {\bibinfo  {journal}
  {Europhys. Lett.}\ }\textbf {\bibinfo {volume} {101}},\ \bibinfo {pages}
  {33001} (\bibinfo {year} {2013})}\BibitemShut {NoStop}%
\bibitem [{\citenamefont {Helbig}\ \emph {et~al.}(2011)\citenamefont {Helbig},
  \citenamefont {Fuks}, \citenamefont {Casula}, \citenamefont {Verstraete},
  \citenamefont {Marques}, \citenamefont {Tokatly},\ and\ \citenamefont
  {Rubio}}]{1dlda}%
  \BibitemOpen
  \bibfield  {author} {\bibinfo {author} {\bibfnamefont {N.}~\bibnamefont
  {Helbig}}, \bibinfo {author} {\bibfnamefont {J.~I.}\ \bibnamefont {Fuks}},
  \bibinfo {author} {\bibfnamefont {M.}~\bibnamefont {Casula}}, \bibinfo
  {author} {\bibfnamefont {M.~J.}\ \bibnamefont {Verstraete}}, \bibinfo
  {author} {\bibfnamefont {M.~A.~L.}\ \bibnamefont {Marques}}, \bibinfo
  {author} {\bibfnamefont {I.~V.}\ \bibnamefont {Tokatly}}, \ and\ \bibinfo
  {author} {\bibfnamefont {A.}~\bibnamefont {Rubio}},\ }\href@noop {}
  {\bibfield  {journal} {\bibinfo  {journal} {Phys. Rev. A}\ }\textbf {\bibinfo
  {volume} {83}},\ \bibinfo {pages} {032503} (\bibinfo {year}
  {2011})}\BibitemShut {NoStop}%
\bibitem [{\citenamefont {D'Agosta}\ and\ \citenamefont
  {Vignale}(2006)}]{AV06}%
  \BibitemOpen
  \bibfield  {author} {\bibinfo {author} {\bibfnamefont {R.}~\bibnamefont
  {D'Agosta}}\ and\ \bibinfo {author} {\bibfnamefont {G.}~\bibnamefont
  {Vignale}},\ }\href@noop {} {\bibfield  {journal} {\bibinfo  {journal} {Phys.
  Rev. Lett.}\ }\textbf {\bibinfo {volume} {96}},\ \bibinfo {pages} {016405}
  (\bibinfo {year} {2006})}\BibitemShut {NoStop}%
\bibitem [{\citenamefont {Buijse}\ \emph {et~al.}(1989)\citenamefont {Buijse},
  \citenamefont {Baerends},\ and\ \citenamefont {Snijders}}]{kine1}%
  \BibitemOpen
  \bibfield  {author} {\bibinfo {author} {\bibfnamefont {M.~A.}\ \bibnamefont
  {Buijse}}, \bibinfo {author} {\bibfnamefont {E.~J.}\ \bibnamefont
  {Baerends}}, \ and\ \bibinfo {author} {\bibfnamefont {J.~G.}\ \bibnamefont
  {Snijders}},\ }\href@noop {} {\bibfield  {journal} {\bibinfo  {journal}
  {Phys. Rev. A}\ }\textbf {\bibinfo {volume} {40}},\ \bibinfo {pages} {4190}
  (\bibinfo {year} {1989})}\BibitemShut {NoStop}%
\bibitem [{\citenamefont {Gritsenko}\ \emph {et~al.}(1996)\citenamefont
  {Gritsenko}, \citenamefont {van Leeuwen},\ and\ \citenamefont
  {Baerends}}]{kine2}%
  \BibitemOpen
  \bibfield  {author} {\bibinfo {author} {\bibfnamefont {O.~V.}\ \bibnamefont
  {Gritsenko}}, \bibinfo {author} {\bibfnamefont {R.}~\bibnamefont {van
  Leeuwen}}, \ and\ \bibinfo {author} {\bibfnamefont {E.~J.}\ \bibnamefont
  {Baerends}},\ }\href@noop {} {\bibfield  {journal} {\bibinfo  {journal} {J.
  Chem. Phys.}\ }\textbf {\bibinfo {volume} {104}},\ \bibinfo {pages} {8535}
  (\bibinfo {year} {1996})}\BibitemShut {NoStop}%
\bibitem [{\citenamefont {Thiele}\ \emph {et~al.}(2008)\citenamefont {Thiele},
  \citenamefont {Gross},\ and\ \citenamefont {K{\"u}mmel}}]{thiele}%
  \BibitemOpen
  \bibfield  {author} {\bibinfo {author} {\bibfnamefont {M.}~\bibnamefont
  {Thiele}}, \bibinfo {author} {\bibfnamefont {E.~K.~U.}\ \bibnamefont
  {Gross}}, \ and\ \bibinfo {author} {\bibfnamefont {S.}~\bibnamefont
  {K{\"u}mmel}},\ }\href@noop {} {\bibfield  {journal} {\bibinfo  {journal}
  {Phys. Rev. Lett.}\ }\textbf {\bibinfo {volume} {100}},\ \bibinfo {pages}
  {153004} (\bibinfo {year} {2008})}\BibitemShut {NoStop}%
\bibitem [{\citenamefont {Taylor}(2006)}]{Taylorbook}%
  \BibitemOpen
  \bibfield  {author} {\bibinfo {author} {\bibfnamefont {J.~R.}\ \bibnamefont
  {Taylor}},\ }\href@noop {} {\emph {\bibinfo {title} {Scattering Theory: The
  Quantum Theory of Nonrelativistic Collisions}}}\ (\bibinfo  {publisher}
  {Dover Publications},\ \bibinfo {year} {2006})\BibitemShut {NoStop}%
\bibitem [{\citenamefont {Fuks}\ \emph {et~al.}(2015)\citenamefont {Fuks},
  \citenamefont {Luo}, \citenamefont {Sandoval},\ and\ \citenamefont
  {Maitra}}]{FLSM15}%
  \BibitemOpen
  \bibfield  {author} {\bibinfo {author} {\bibfnamefont {J.~I.}\ \bibnamefont
  {Fuks}}, \bibinfo {author} {\bibfnamefont {K.}~\bibnamefont {Luo}}, \bibinfo
  {author} {\bibfnamefont {E.~D.}\ \bibnamefont {Sandoval}}, \ and\ \bibinfo
  {author} {\bibfnamefont {N.~T.}\ \bibnamefont {Maitra}},\ }\href@noop {}
  {\bibfield  {journal} {\bibinfo  {journal} {Phys. Rev. Lett.}\ }\textbf
  {\bibinfo {volume} {114}},\ \bibinfo {pages} {183002} (\bibinfo {year}
  {2015})}\BibitemShut {NoStop}%
\end{thebibliography}%

\end{document}